\documentclass[a4paper]{sbbdshort}

\usepackage[T1]{fontenc}   
\usepackage{graphicx}
\usepackage{epsfig}

\usepackage[english]{babel}
\usepackage[utf8]{inputenc}

\usepackage{subfig}
\graphicspath{{./}}
\usepackage{verbatim}

\makeatletter
\newif\if@restonecol
\makeatother

\usepackage[ruled,vlined]{algorithm2e}
\usepackage{listings}
\usepackage{xcolor}
\usepackage{graphicx}

\lstnewenvironment{java}{\lstset{language=Java, basicstyle=\color{black}\footnotesize,frame=tb}}{}

\newdef{definition}[theorem]{Definition}
\newdef{remark}[theorem]{Remark}

\makeatletter
\def\thepage{130-\number\numexpr\c@page\relax}
\makeatother

\title{Testing MapReduce-Based Systems}

\author{João Eugenio Marynowski\inst{1}, Michel Albonico\inst{1}, Eduardo Cunha de Almeida\inst{1}, Gerson Sunyé\inst{2}}

\institute{Federal University of Paraná, Brazil \\ \email{\{jeugenio,albonico,eduardo\}@inf.ufpr.br}
\and INRIA - University of Nantes, France \\ \email{gerson.sunye@univ-nantes.fr}
}

\begin{abstract}
MapReduce (MR) is the most popular solution to build applications for large-scale data processing.
These applications are often deployed on large clusters of commodity machines, where failures happen constantly due to bugs, hardware problems, and outages.
Testing MR-based systems is hard, since it is needed a great effort of test harness to execute distributed test cases upon failures.
In this paper, we present a novel testing solution to tackle this issue called HadoopTest.
This solution is based on a scalable harness approach, where distributed tester components are hung around each map and reduce worker (i.e., node). 
Testers are allowed to stimulate each worker to inject failures on them, monitor their behavior, and validate testing results.
HadoopTest was used to test two applications bundled into Hadoop, the Apache open source MapReduce implementation.
Our initial implementation demonstrates promising results, with HadoopTest coordinating  test cases across distributed MapReduce workers, and finding bugs.
\end{abstract}



\begin{document}

\begin{bottomstuff}
Work partially funded by the Datluge CNPq-INRIA project.
\end{bottomstuff}

\maketitle

\section{Introduction}

Over the last decade the amount of data generated by several applications, like customer reports, survey data and social networks, have reached several petabytes and this amount of data tends to increase dramatically along the next few years up to iotabytes~\cite{DBLP:conf/icde/WinterK10}.
The analysis of this large amount of data requires a great effort of data processing, drawing a widespread attention from both, the academic community and the industry.
Thus, new types of frameworks for large-scale data processing have been proposed. One of them, MapReduce (MR)~\cite{DBLP:conf/osdi/DeanG04} appears as the most popular one. It allows non-expert users to easily use a large number of machines to process a large amount of data. 
Among several implementations of MapReduce, the open-source framework Hadoop~\cite{Hadoop} from the Apache Foundation has been used by a growing number of companies, such as: EBay, Linkdln, Quantcast, Facebook and Yahoo!~\cite{HadoopUsers}.
Moreover, some DBMS vendors, including Aster, Greenplum and Vertica, have started to integrate MR front-ends into their systems to run on large-cluster machines.

As any other large-scale environment, MR implementations constantly face failures due to several different conditions (e.g., outages, hardware problems, bugs)~\cite{yunhao:2004a}. Thus, these implementations must be designed to be fault-tolerant and tested intensively to ensure they are reliable. However, testing large-scale systems is a complex task and requires a test harness, i.e. a framework to build tests and control their execution by monitoring the behavior of a system under test (SUT) and its outputs~\cite{walter:1998}.
The main issue of a test harness is to control the execution of distributed test cases across a large-scale environment. 
More precisely, this control requires:  the coordination of distributed test cases, the monitoring of the nodes that execute the test cases, and the injection of faults to reproduce real-world conditions.
However, available solutions for testing MR implementations are limited and none of them tackle completely this issue.

In this work, we present HadoopTest, a test harness for MR-based implementations. HadoopTest deploys and manages the execution of test cases across distributed nodes.
This is achieved by controlling each MR worker (i.e. node) with a distributed \emph{Tester}.
A tester  is a component that controls the overall test case execution through a distributed coordination algorithm.
HadoopTest was used to test two applications bundled with Hadoop. 
The initial implementation demonstrates promising results, with HadoopTest coordinating  test cases across distributed MR workers and finding bugs.

The rest of the paper is organized as follows.
The next session introduces the basic concepts of MapReduce and software testing. 
Section~\ref{sec:hadooptest} presents our framework for testing MR systems.
Section~\ref{sec:experiments} describes the initial results through implementation and experimentation.
Section~\ref{sec:related} discusses related work.
Section~\ref{sec:conclusion} concludes.

%
\section{Basic Concepts}\label{sec:basic}
\begin{figure}[htbp]
	\centering
	\includegraphics[scale=0.6]{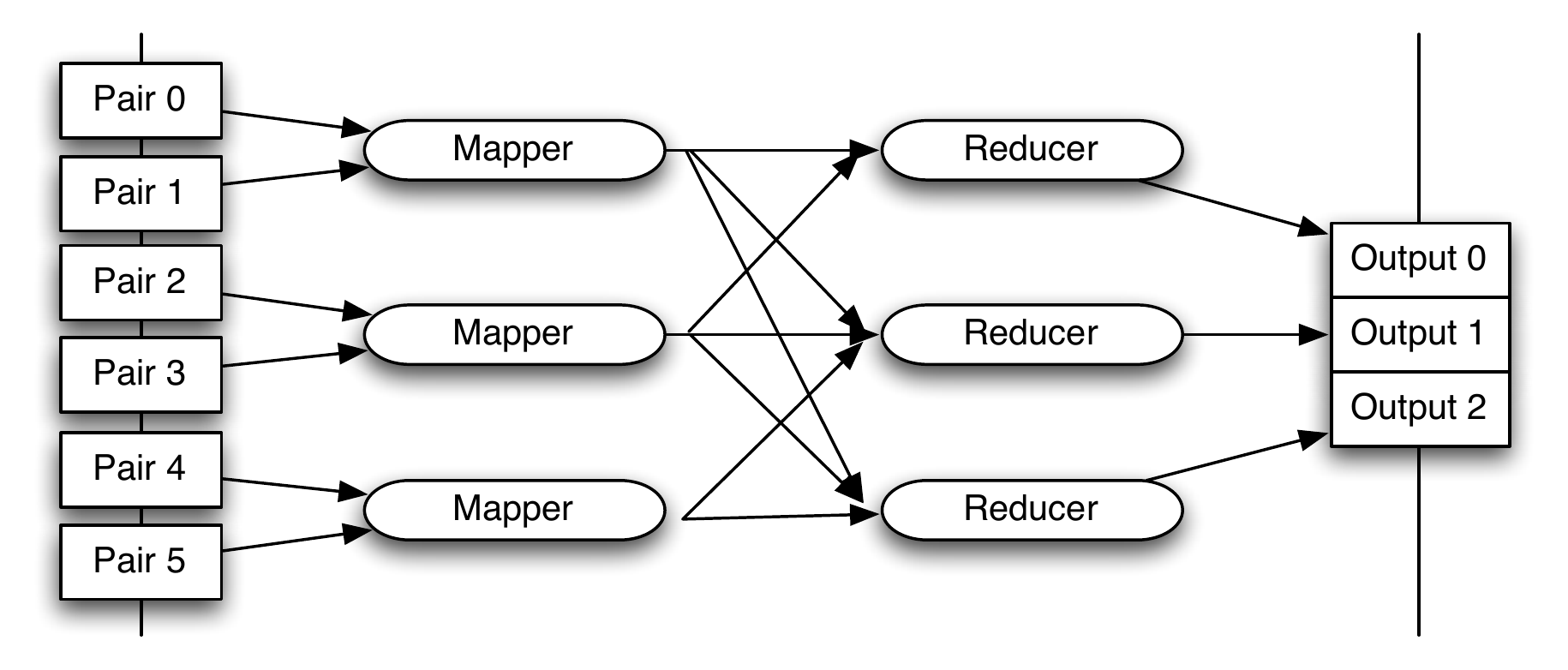}
	\caption{Execution of Map and Reduce operations}\label{fig:mrexecute}
\end{figure}

MapReduce has a simplified programming model, leaving aside the treatment of events related to the distributed environment, without programmer intervention.
Its programming environment is based on two higher-order functions: Map and Reduce.
Each one of these functions has a well-defined behavior.
They work together, where the mapping is the initial analysis of the data input and the output is handled by the reduce function.
Both the map and the reduce functions are programmed by the user.
Several copies of the program are distributed across machine-nodes, where one copy is called master and the other copies are workers.
The master selects idle workers to assign a map or a reduce instance.  
The data flow  between the map and the reduce functions  is shown in Figure~\ref{fig:mrexecute}.

All processing is done based on $<key,value>$ pairs.
For each key, the  map function associates a value.
The user  defines the correspondence between the \textit{value} and \textit{key}.
The output of the map function is said to be the intermediate key pair.

For each range of keys will be assigned an intermediate \textit{Reducer}, a service responsible for performing the reduce function.
As a result, the operation will generate a pair $<key, result>$.
Analogous to the map function, it is up to the user to define the correspondence between the \textit{key} and result.
When the map function terminates, the reduce function starts.
When all the reduce instances terminate, they append their result to a final output file.

\subsection{Testing Background}

Testing is a process to validate a system to ensure its quality~\cite{_AMMANN_}.
In our experiments, we focus on functional testing, whose principle is to apply inputs to the SUT and to compare the observed outputs to the expected results.
This comparison is made by an \textit{oracle}, the mechanism responsible for assigning a \textit{verdict}.
If the expected results and the observed outputs are the same, the verdict is \textit{pass}.
Otherwise, the verdict is \textit{fail}.
The verdict may also be \textit{inconclusive}, meaning that the test result is not precise enough to satisfy the test criteria.

There are different sorts of oracles~\cite{baresi:2001b}, for instance, assertions, value comparison, log file analysis, and manual analysis. 
Our approach is based on assertions and value comparison due to their simplicity. 
Both types of test oracles are sufficient to the HadoopTest initial implementation, since we are testing simple MR applications.
However, we plan to add a broader range of test oracles in the future.

\section{Testing MapReduce-based Systems}
\label{sec:hadooptest}

Testing MR-based systems is a complex task, for which the test harness must execute distributed test cases.
The main issues of the testing harness are: the  coordination of distributed test cases,  the control of distributed workers and the injection of faults.

Distributed testing is only possible through a mechanism to coordinate the execution of distributed test cases across distributed nodes.
In the context of MR, this can be more complex, since several workers with different functionalities are running together.
For instance, while a worker is applying a map function another is leaving the system due to a fault.
This also leads to an individual control of the workers, which is important to put the workers in any state along testing.

The fault injection issue is important to reproduce a real-world environment during testing. 
Several types of failures must be considered, such as: node failure, network traffic, high latency and performance variations.
In addition, these faults may be executed individually at each worker.
For instance, to reject the disk access to a specific reduce instance to test fault-tolerance.

%
\subsection{HadoopTest}

In this section, we present HadoopTest, a test harness solution for MR-based systems.
This solution is based on the individual control of distributed MR workers across large-cluster machines.
This approach allows to coordinate the execution of test case actions on different workers in parallel.
HadoopTest allows the combination of fault injection and functional tests to build complex test cases.
Test cases deployed on HadoopTest are not intrusive, meaning that API calls are not added to the SUT source code. 
The reason is that additional code may add bugs to the SUT, making it difficult to detect the source of a failure, which could be either the SUT code or the inserted code.

Test cases are written in Java, similarly to JUnit~\cite{junit}, to accelerate HadoopTest acceptance, although we plan to improve the testing language in the future (e.g., using Bloom~\cite{DBLP:conf/eurosys/AlvaroCCEHS10}). 
More precisely, test cases are Java classes and test case actions are instance methods, marked with an annotation. 
Annotations are metadata where test engineers define deployment instructions and the test case execution workflow.   
HadoopTest uses reflection to read this metadata to achieve coordinating the test cases (detailed in Section~\ref{sec:synch}).

\begin{figure}[h]
  \centering
  \subfloat[Deployment diagram]{
    \includegraphics[width=6.5cm]{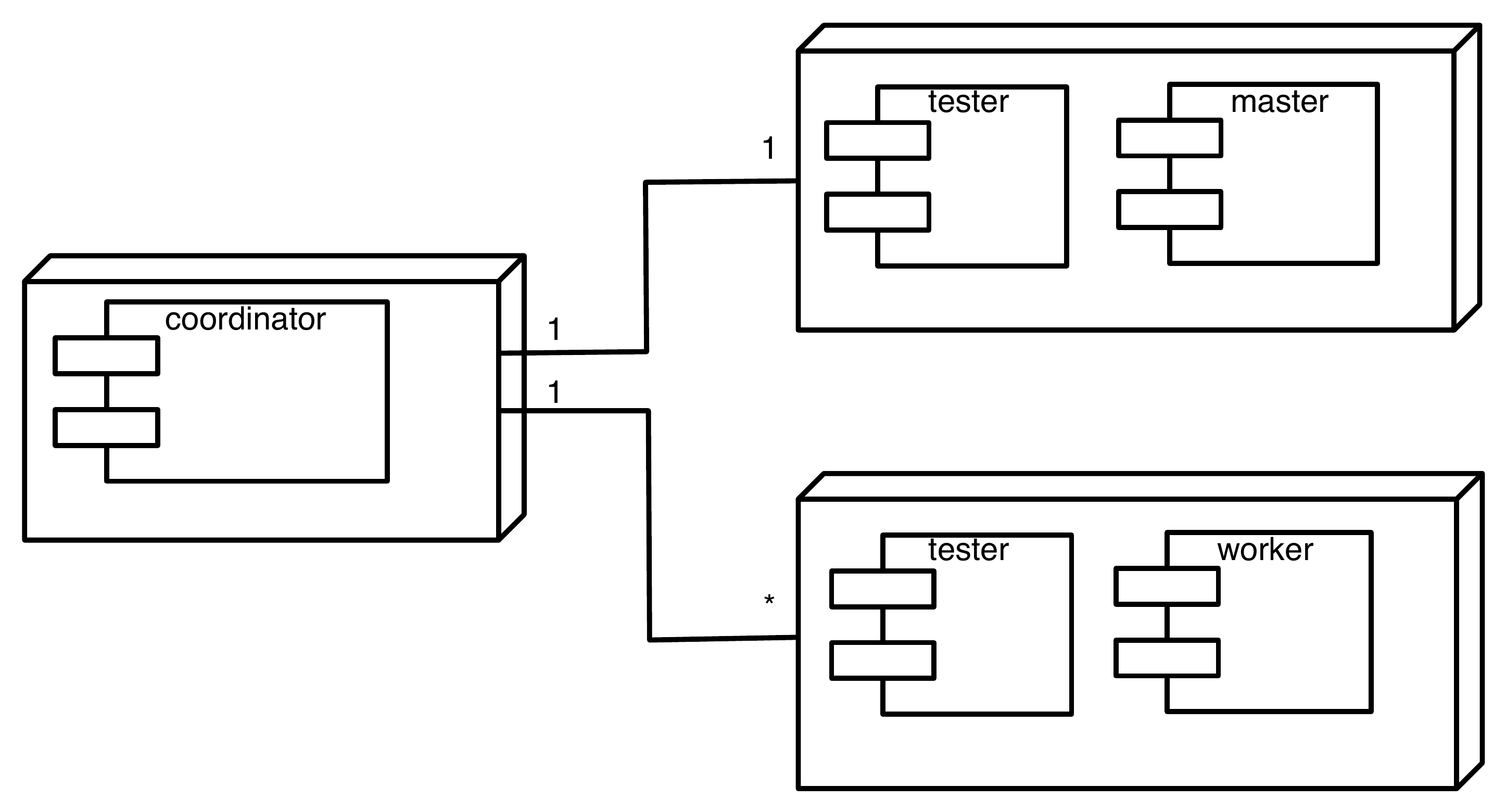}
    \label{fig:coordtestmr}
  }
  \hfill
  \subfloat[Testing a MapReduce instance]{
    \includegraphics[width=8cm]{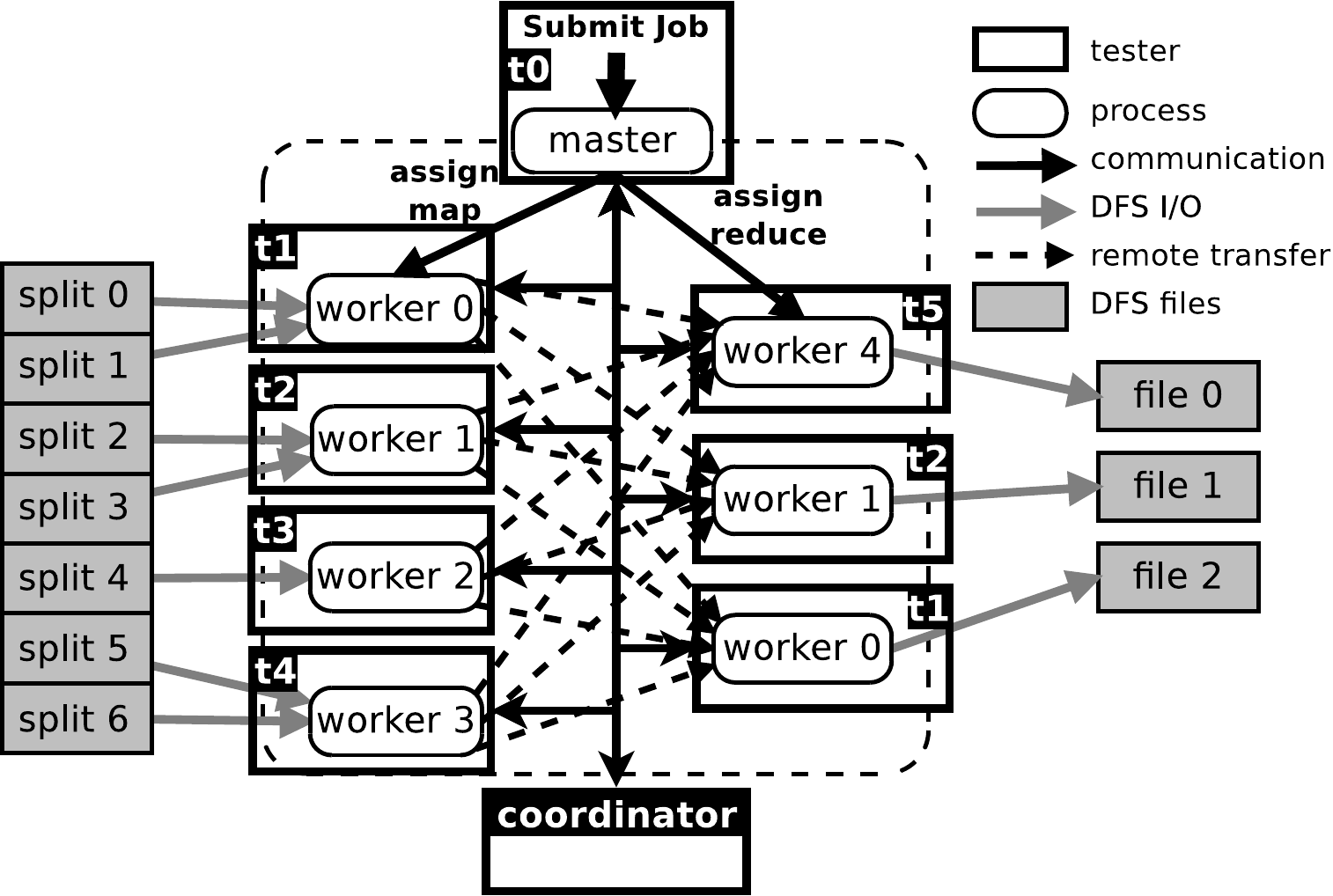}
    \label{fig:MRTesting}
  }
  \caption{HadoopTest diagrams}
\end{figure}

The HadoopTest architecture, shown in Figure~\ref{fig:coordtestmr}, is composed of a test \textit{coordinator} and distributed \textit{testers}.
The coordinator is responsible for three main tasks: (1) to control the execution of distributed test cases, (2) to coordinate the test case actions and (3) to gather the verdicts from the testers.
The testers receive coordination messages from the \textit{coordinator} to execute the test case actions on the master and distributed workers.
These messages can also contain stimulus to inject faults on the workers. 
In the present version of HadoopTest, we only implemented outages to drop workers. Other types of failures depend on a lower-level control of the SUT, for instance, disk or network failures. 
Once the testers reach the end of a test case execution, they validate the results to assign a local verdict. 
This validation is based on assertions and value comparison.

Figure~\ref{fig:MRTesting} shows the application of HadoopTest to an MR instance.
The \textit{coordinator} controls the execution of six testers, identified by $t0..t5$.
Tester $t0$ controls the master and each one of the other testers, $t1..t5$, controls a worker instance.


\subsubsection{The Coordination of Actions}\label{sec:synch}
We let  $\tau = (A^\tau, V^\tau, T^\tau)$ denote a distributed test case,  where $A^\tau$ is a set of test actions, $V^\tau$ is a set of local verdicts, and  $T^\tau$ is a set of testers.
A test case action is a tuple $a = (T', I, \theta, h)$ where $T'$ is a set of testers that should execute the action, $I$ is a set of instructions (e.g., Java instructions), and $\theta$ is the interval of time in which the action should be executed.
This interval is important to bound the execution time, since coordination messages are asynchronous.
We let $h$ denote the hierarchical order that the actions will execute.
Naturally, actions with the same hierarchical order are allowed to execute in parallel.  

The overall coordination approach is outlined in Algorithm~\ref{alg:test_exec}.
Initially, the coordinator registers into a schedule $S^\tau$ each tester $t$ and the test actions $A^{\tau}$ that such tester will invoke. 
The schedule $S^\tau$ maps actions onto sets of testers. 
Once the registration is finished, one coordination message is sent to each corresponding tester following the sequence of actions.    
The coordination messages are sent asynchronously to allow the execution of the actions in parallel.
Local verdicts are also received asynchronously from the testers.
If an action takes more than $\theta$, then the verdict is \textit{inconclusive}.
At the end, the oracle returns the final verdict that gathers all the local verdicts.

\begin{algorithm}
\caption{Coordination Algorithm}
\label{alg:test_exec}
\KwIn{ $A^\tau$, an ordered set of test actions; $T^\tau$, a set of testers;}
\KwOut{ a global verdict}
{
	$S^\tau \gets \emptyset$\; $V^\tau \gets \emptyset$\; 
	\ForAll{$t \in T^\tau$}{
		$S^\tau \gets register(t, A^{\tau})$\;
	}
		\ForAll{$a \in A^\tau$ such that $h$ is the same}{
			Send coordination messages for all $t \in T'$ such that $a \mapsto T'$ up to $\theta$\;
			\If{$\theta$}{
				$V^\tau \gets inconclusive$\;
			}	
			\Else{
				$V^\tau \gets t.result(a^{h})$\;
			}	
		}
	\Return{$oracle(V^\tau)$} \;
}
\end{algorithm}

\subsubsection{Test Case Example}


Table~\ref{tab:MRTestCase} shows a simple distributed test case to illustrate the algorithm.
The goal is to validate a computation upon failures.
This test case involves six testers $T = \{t_0, \ldots, t_5\}$ and seven test actions $A = \{a_0, \ldots, a_6\}$. 
The tester $t_0$ executes the action $a_0$ to start the master.
The testers $\{t_1, \ldots, t_5\}$ execute the action $a_1$ to start the workers.
Next, tester $t_0$ submits a job to the master at action $a_2$.
This job is composed by a computation in the form of map and reduce functions.
During the execution of the job, the tester $t_2$ is dropped from the system.
The validation is done at action $a_4$.
If the output data is the same as the expected, then the verdict is $pass$. 
Otherwise,  the verdict is $fail$. 
If $t_0$ is not able to retrieve any output data, then the verdict is $inconclusive$.
The rest of the actions are executed to stop the SUT.

\begin{table}[h]
\begin{center}
\caption{MapReduce simple test case}
\label{tab:MRTestCase}
\begin{tabular}{|c|c|l|c|l|}
\hline 
Action & Hierarchy & Tester	& Instructions  & Timeout \\
\hline\hline
$a_0$ & $0$ & $t_0$	    & startMaster() &  100 \\
$a_1$ & $1$ & $t_1$-$t_5$	& startWorkers() &   1000 \\
$a_2$ & $2$ & $t_0$     & sendJob() & 1000000 \\
$a_3$ & $2$	& $t_2$	    & dropWorker() & 1000 \\
$a_4$ & $3$ & $t_0$	    & assert(output, expected) &  10000 \\
$a_5$ & $4$ & $t_1,t_3$-$t_5$	& stopWorkers() & 1000 \\
$a_6$ & $5$ & $t_0$	    & stopMaster() & 1000 \\
\hline
\end{tabular}
\end{center}
\end{table}

\section{Experimental Validation}
\label{sec:experiments}

This section presents the evaluation of HadoopTest through experimentation.
First, we evaluated the overhead produced by HadoopTest to coordinate the execution of distributed test cases.
Second, we used HadoopTest to test two applications bundled into Hadoop, however, due to the lack of space, we only present the PiEstimator results. The objective is to validate whether HadoopTest is able to identify bugs.

The PiEstimator calculates the $\pi$ value using the Monte Carlo method, that considers a circle exactly inscribed inside a square with side length $1$.
The map function randomly creates points inside the square and counts the points placed inside and outside the circle.
The reduce function accumulates the points inside ($I$) and outside ($O$) counted by the map function.
The estimated $\pi$ value is obtained by $4*(I/T)$, were $T=I+O$.

The experiments were executed on the Grid5000 platform\footnote{Grid 5000 Platform: http://www.grid5000.fr} using up to 50 cluster machines running Debian GNU/Linux. The cluster machines were connected by a 1 Gbps network and they had a similar configuration: 2 Intel Xeon 2.6GHz dual-core processors, 8 GB RAM memory and 250 GB SATA HD.

\subsection{The HadoopTest Overhead}
\label{sec:integrity}

\begin{figure}[h]
	\centering
	\includegraphics[width=.7\textwidth]{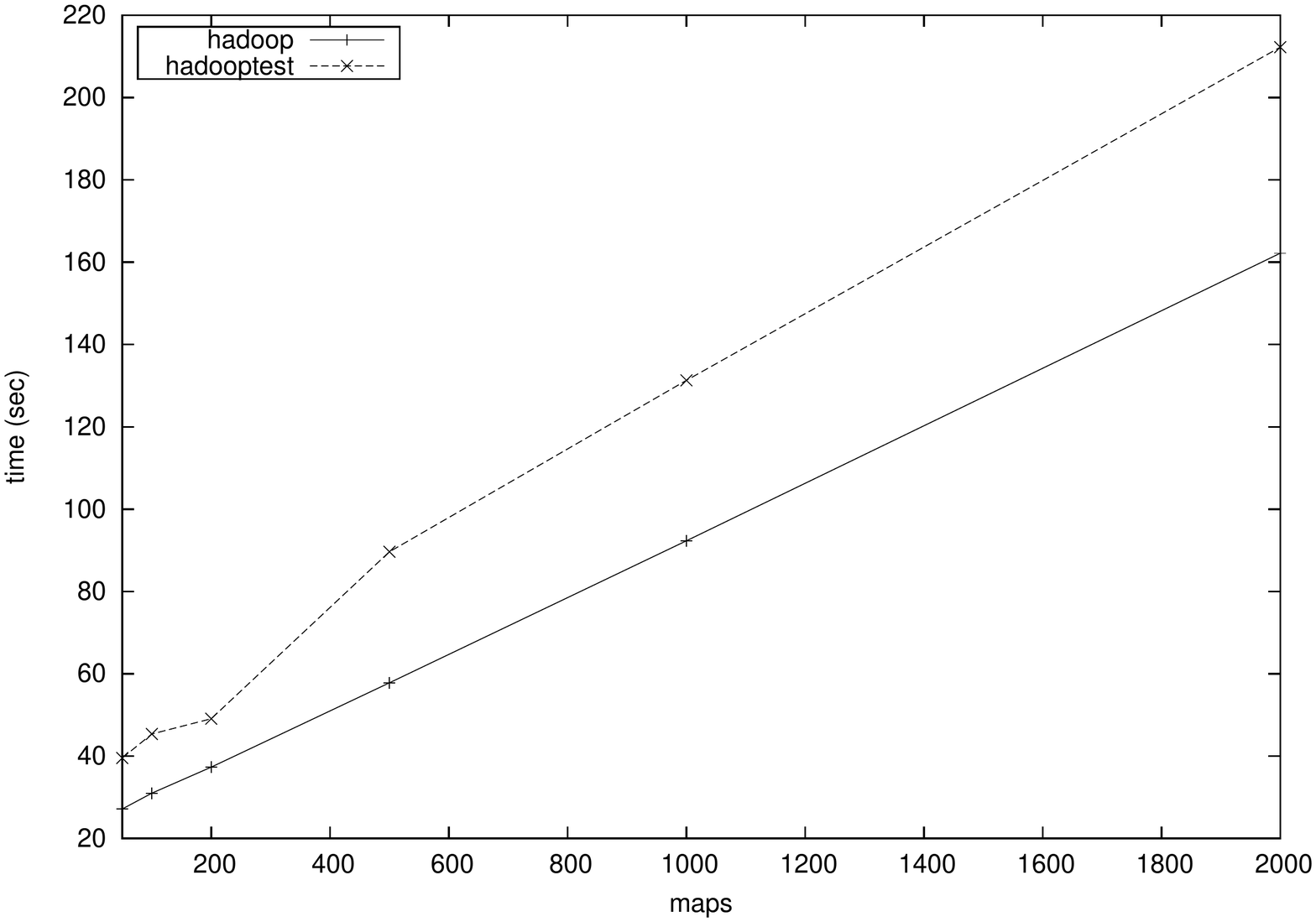}
	\caption{Execution time variance of the PiEstimator}
	\label{fig:hadoopXhadooptest}
\end{figure}

To evaluate the overhead  produced by HadoopTest, we executed the PiEstimator application in two ways. 
In the first one, Hadoop is executed alone, in order to evaluate the raw execution time.
In the second one, Hadoop is executed along with HadoopTest, in order to evaluate the overhead produced during testing.
We perform tests with 2, 10, 20, and 50 machine-nodes on the Grid5000.
Figure~\ref{fig:hadoopXhadooptest} shows the execution time of PiEstimator running on Hadoop and HadoopTest with 50 machine-nodes.
We vary the number of map instances in each execution.

The execution with HadoopTest presented an average overhead of 30\%.
Although this overhead decreases the performance of the tests, it did not harm the test results.
In fact, this overhead is important while testing applications where timing constraints are important.
Future work is needed to reduce the coordination messages and consequently, reducing the overhead. 
Nevertheless, HadoopTest coordinated up to 10,000 map instances concurrently, however, the overhead due to the number of threads running along was overwhelming to 50  machine-nodes. 
A larger number of machine-nodes is required to test this number of map instances.

\subsection{The HadoopTest Results}
\label{sec:mutants}
To evaluate whether HadoopTest is able to identify bugs, we used Mutation Analysis~\cite{ma:2006,DBLP:conf/itc/Offutt94} to create a set of buggy versions (i.e., mutants) of the PiEstimator. We used the ASM~\cite{bruneton:2002} Java bytecode engineering library to create mutants.
In our experiments, a mutation is a modification of an arithmetic and/or a logic operator into the original bytecode to generate an incorrect result. The goal is to identify the largest possible number of incorrect results.

We generated 13 mutants of the PiEstimator class and the results are shown in Table II.
The expected $\pi$ value returned by the original application was $3.1416$ and only the mutants M1, M6, M7, M9 and M12 returned this value.
These mutants have the \textit{Pass} verdict on the test case execution while the other mutants M4, M5, M10 and M11 received a \textit{Fail} verdict, since the $\pi$ computation parameters were modified resulting in a different value than the expected one.
In the case of the mutants M0, M2, M3 and M8, the modifications were in the execution parameters which interfered on their correct execution. 
They returned NULL as results. 

\begin{table}[htdp]
\caption{Test case over 13 mutants from the PiEstimator}
\begin{center}
\renewcommand{\tabcolsep}{0.55cm}
\begin{tabular}{|l|l|c|c|}
\hline 
\textbf{Mutants} & \textbf{Result} & \textbf{Pass} &  \textbf{Fail} \\
\hline\hline
M0 & NULL & & X \\
\hline
M1 & 3.1416 & X & \\
\hline
M2 & NULL &  & X \\
\hline
M3 & NULL &  & X \\
\hline
M4 & 3.0776 &  & X \\
\hline
M5 & 3.1312   &  & X \\
\hline
M6 & 3.1416  & X & \\
\hline
M7 & 3.1416 & X & \\
\hline
M8 & NULL &  & X \\
\hline
M9 & 3.1416 & X & \\
\hline
M10 & 3.1408 &  & X \\
\hline
M11 & 3.1408 &  & X \\ 
\hline
M12 & 3.1416 & X &  \\
\hline
\end{tabular}
\end{center}
\label{tab:PiResults}
\end{table}

HadoopTest effectiveness was evaluated in terms of the number of detected mutants. 
When mutation analysis is applied to a program code and generates several mutants, some of them are equivalents to the original source for different reasons: the modified part is never executed, binary operators are equivalents for this precise case, etc. 
Here, we considered that the mutants that obtained the same output as the original application are equivalents.
The initial implementation of HadoopTest demonstrated promising results by identifying all of the non-equivalent mutants of the PiEstimator.
This was also true for the WordCount application tests.
\section{Related Work}
\label{sec:related}

MRUnit~\cite{mrunit:2010} and Herriot~\cite{herriot:2010} are the most popular tools to test MR implementations.
They are designed to help bridging MR implementations and JUnit testing tool.
Both tools provide an API that must be used along the development of an MR-based system.
Differently from MRUnit, Herriot provides an external component, Test Node, which is responsible for the overall test execution control.
This control is only possible including some API calls into the SUT source-code.
However, this intrusive approach may spoil the source code generating new bugs.
Differently from our approach, both tools neither coordinate different nodes at the same time nor stimulate them in parallel which restricts to build complex test cases.

Ganesha~\cite{DBLP:journals/sigmetrics/PanTKGN09} is a black-box testing tool used to detect performance problems in MR systems.
Differently from MRUnit and Herriot, it does not require any modification on the SUT source code.
Ganesha differs from our approach in the validation aspect. 
It monitors both, the OS-level and Network-level counters to detect the culprit node that compromised the MR performance.
This approach produces low performance overhead, however, the validation is narrowed by the limitations of the OS-level and Network-level counters. In addition,  Ganesha requires an additional effort to filter the metrics of concurrent applications out of the counters (including MR's).
In contrast, HadoopTest provides on-line assertion mechanisms that can be used by any kind of test (including performance test) as well as the individual control of nodes, which can be used to detect any performance problem.

PeerUnit~\cite{DBLP:journals/ese/AlmeidaSTV10,DBLP:conf/kbse/AlmeidaMSV10} is a framework for testing peer-to-peer systems.
It ensures the sequencing of test case actions synchronizing their execution through distributed nodes. 
PeerUnit allows to coordinate any SUT component also leveraging the approach of distributed testers.
It differs from our approach on the test case execution workflow.
First, PeerUnit can not execute different actions in parallel, preventing the writing of complex test cases.
Second, nodes (i.e., peers) are considered functionally identical, which is not the case of MR implementations, where they are divided into: workers and master nodes (as detailed in Section~\ref{sec:basic}).  
Moreover, different workers may perform different actions along the execution  of an MR application.
In our approach, distributed testers are designed to achieve handling both nodes and actions in parallel.

\section{Conclusion}
\label{sec:conclusion}

In this paper we presented HadoopTest,  a test harness solution for testing MapReduce-based systems.
HadoopTest tackles two important issues of MapReduce testing.
First, it coordinates the execution of test cases across distributed  MapReduce workers.
This is achieved by controlling each MR worker with distributed testers.
Second, HadoopTest allows the coordination of the parallel execution of test case actions on different MR workers.
Naturally, in practice it allows to combine fault injection with functional tests to built complex test cases. 

Through experimentation on 50 machine-nodes on the Grid5000, we showed that HadoopTest coordinates distributed test cases across MR workers.
We also used HadoopTest to test two applications bundled into Hadoop,  the PiEstimator and  the WordCount.
We used mutation analysis to create mutants of these applications.
HadoopTest demonstrated promising results by identifying all of the non-equivalent mutants.

In the future, we plan to execute more complex experiments involving different fault injections. This can be achieved by instrumenting the underlying structure of the SUT (e.g., network, disk), which is not yet possible with this version of HadoopTest.
In addition, we plan to test more complex MapReduce applications, such as: Hive, Mahout, and Nutch.
These applications provide a broader range of functionalities to test. 

\bibliographystyle{jidm}
\bibliography{sbbd2011short}

\begin{received}
\end{received}

\end{document}